\documentstyle[12pt]{article}
\begin{document}\setcounter{page}{0}
\def\R{{\cal R}}
\setlength{\unitlength}{1mm}

\title{Complete angular analysis of the  decay cascade $B \rightarrow
D^{**}( \rightarrow D^*(\rightarrow D \pi) + \pi) + W(\rightarrow l  \nu)$}

\author{
P. Bia\l as\thanks{Supported in part by the KBN grant 203809101}
\thanks{e-mail address: pbialas@vsk02.ifj.edu.pl},
K. Zalewski$^*\,$\thanks{also at Institut of Nuclear Physics, Krak\'ow}
\,\thanks{e-mail address: zalewski@vsk02.ifj.edu.pl}
\\
Institute of Physics\\
Jagellonian University\\
30-059 Krak\'ow, ul. Reymonta 4, Poland\\[0.5cm]
J.G. K\"orner\thanks{Supported in part by the BMFT, FRG under contract
 06MZ760}\\
Institut f\"ur Physik,\\
Johannes Gutenberg-Universit\"at,\\
Staudinger Weg 7,
Postfach 3980 \\
D-6500 Mainz,
Germany \\[0.5cm]
}
\date{}

\maketitle

\begin{abstract}

We develop a general algorithm for describing angular decay
distributions in cascade decay chains of arbitrary length. The general
algorithm is used to study
joint angular decay distributions for the cascade decay $B
\rightarrow D^{**}( \rightarrow D^*(\rightarrow D \pi)+\pi) + W(\rightarrow
l\nu)$ where the $D^{**}$ is a generic $P$-wave charm meson state.
Lepton mass effects are fully incorporated
The joint angular decay distribution
depends on 43 independently measurable decay parameters if the spin
parity of the $D^{**}$ is $1^+$ and on 48 decay parameters if the
spin parity of the $D^{**}$ is $2^+$.  We give expressions for these
decay parameters in terms of the helicity amplitudes of the two-body
 decay processes. An absolute prediction for all the
parameters  is
presented in the framework of the heavy quark effective theory. A method for
obtaining the helicity amplitudes from
measured joint angular distributions is suggested.
\end{abstract}
\vfill

\noindent TPJU-19/92\\
MZ-TH/92-53\\
December 1992\\
\newpage

\section{Introduction}

Semileptonic decays of $B$ mesons have been much studied both experimentally
and theoretically. For recent reviews cf. e.g. \cite{STO} for experiment and
\cite{NAR} for theory. On the experimental side ARGUS \cite{ARG} and
CLEO \cite{CL1,CL2} find that
the ground state to ground state
decays\footnote{Here and in the following when bars over the
antiparticles are omitted, we mean the pair of charge conjugated
decays} $B \rightarrow Dl\nu$ and $B \rightarrow D^* l \nu$ do not
saturate the total semileptonic decay width. For the contribution of
the ground state to excited state transitions
 ARGUS finds ($40 \pm 10$) per cent
\cite{ARG} and CLEO has ($32 \pm 5$) per cent \cite{CL2}. Not much is
known experimentally about the nature of these other channels. It is,
however, natural to expect that the $P$-wave resonances $D^{**}$
should play an important role in them. The Particle Data Group
\cite{PDG} lists  two $P$-wave candidates,
the $2^+$ resonance $D_2(2460)$ and the $1^+$
resonance $D_1(2420)$. They can be plausibly interpreted
as composed of a $c$ quark almost at rest in the meson rest frame and
a $P$-wave light antiquark with total angular momentum $j = 3/2$ \cite{LWI}.
Moreover, one expects \cite{LWI} two more $D^{**}$ resonances ($0^+$,
$1^+$) with the $P$ wave antiquark in the $j=1/2$ state.

The theory of the reactions $B \rightarrow D^{**}(\rightarrow
\ldots)+W(\rightarrow l\nu)$ and of the
subsequent $D^{**}$ decays has been discussed by
a number of authors
%\cite{LWI},\cite{ISG}, \cite{IW2},
%\cite{IW3}, \cite{KZ0} \cite{CNP}, \cite{BKT} \cite{KKP}
[7-14]. From angular
momentum and parity conservation the allowed $D^{**}$ decays are
$D^{**}(0^+) \rightarrow (D+\pi)_S$, $D^{**}(1^+) \rightarrow
(D^*\pi)_{S or D}$, $D^{**}(2^+) \rightarrow (D^*\pi)_D$, and
$D^{**}(2^+) \rightarrow (D\pi)_D$, where the subscript denotes the
partial wave of the decay channel. In the limit $m_c \rightarrow \infty$ one
finds that the $D^{**}(1^+, j=\frac{3}{2})$ decays only into
$(D^*\pi)_D$ and the $D^{**}(1^+, j=\frac{1}{2})$ only into
$(D^*\pi)_S$ \cite{LWI}.  This follows from angular momentum conservation in
the
decay of the $P$-wave antiquark into an $S$-wave antiquark and a pion.
Moreover, for the $D^{**}(2^+)$ decays  the
ratio of probabilities for the channels $D^*\pi$ and $D\pi$ is 3:2 in this
limit \cite{LWI,BKT,KKP}.

Testing how well these predictions apply to mesons containing finite
mass $c$-quarks is very interesting, but difficult for a variety of
reasons. In particular, on the theoretical side,  one has to introduce somewhat
arbitrary kinematical
corrections (cf. e.g. \cite{LWI}) when comparing
different decays. The heavy quark symmetry
predictions for the decay distributions in a given channel are more
reliable than a comparison of different decays. For example, the
predictions of the nonrelativistic quark model
have
been amazingly successful
when they were applied to a given decay channel.
However, the angular distribution for the
decays $D^{**}\rightarrow D^*\pi$ does not contain enough
information by itself. Therefore, e.g.  Ming-Lu, Wise and
Isgur \cite{LWI} proposed to study in addition to standard decay
distributions also the decays into longitudinally polarized $D^*$'s.

We generalize this approach to a complete analysis of the angular
decay distributions for the full cascade decay

\begin{equation}
\label{DEC}
B \rightarrow D^{**}(\rightarrow D^*(\rightarrow D\pi)+\pi)+W(\rightarrow
l\nu)
\end{equation}
These angular decay distributions depend on three pairs of spherical
angles defined by: (i) the $D^*$ in the $D^{**}$ rest frame, (ii) the
$D$ in the $D^*$ rest frame and (iii) the lepton in the rest frame of
the virtual $W$-boson. Other popular distributions, like the energy
distributions for the lepton in the $B$ rest frame or angular decay
distributions for the $D^*$ in special spin states can be expressed
 by these basic distributions (for given masses of the intermediate
particles\footnote{Here and in
the following particle stands for a  particle, a resonance or
the virtual $W$-boson.}). The angular distribution of the $D^{**}$ in
the $B$ rest frame is of no interest, since it has to be spherically
symmetric by angular momentum conservation.
Lepton mass effects are fully incorporated in our approach.
Thus our results are applicable also
for the decays involving the $\tau$-lepton.

The angular decay distributions contain many independently measurable
parameters: 43 for each of the two $D^{**}(1^+)$ cases and 48 for the
$D^{**}(2^+)$ case. All these parameters can be expressed in a model
independent way by the helicity amplitudes. The overall normalization
and the overall phase of the helicity amplitudes drop out and this
reduces the number of real parameters to eight in each of the $1^+$
cases and to six in the $2^+$ case.
When we say that helicity amplitudes are measured or predicted this always
implies a determination up to an overall constant complex factor.
The usual assumption that all the
helicity amplitudes are relatively real (cf. \cite{BKK} and references
quoted there) reduces the numbers of parameters to four (three) for the
decays involving each $D^{**}(1^+)$ ($D^{**}(2^+)$). Each eliminated
phase leaves, however, a sign ambiguity.  A further reduction of the
number of parameters requires a dynamical theory or model. We present
 absolute predictions for the angular decay distributions
(no free parameters) from  Heavy Quark
Effective Theory (HQET) described in \cite{LWI,IW2,IW3,KZ0,BKT}.

The plan of this paper is as follows. In Section 2 we write down
general formulae for  decay distributions following the classical
method of Jacob and Wick \cite{JW}. We introduce a diagrammatic
representation of the decay formulae in order to keep track of the
numerous terms. In Section 3 we use statistical tensors (see
\cite{BKK} and references quoted there)  to  simplify the
formulae and their diagrammatic representation. The discussion in
Section 2 and Section 3 is quite general and applies to simple decays
as well as to cascade decay chains of arbitrary length. In Section 4
we apply our algorithm to the decays in eq. (\ref{DEC}). As an
illustration we show how these formulae
give an absolute prediction for all angular distributions
when supplemented by the predictions of HQET.
In Section 5 we discuss some problems relevant to the
practical analysis of data. Section 6 contains our conclusions.

\section{Diagrammatic representation of decay formulae}

Writing down the joint angular distributions for the decays
in eq. (\ref{DEC}) is in principle very simple. In practice, however,
it is cumbersome to
keep track of the six spin reference frames and of the many
summations. Therefore, we propose a diagrammatic method
which eliminates most of the effort. In the present section we
introduce our method by applying it to the (extended) approach of
Jacob and Wick \cite{JW}.

Let us generalize the decay process in eq. (\ref{DEC}) as follows. A
parent particle (e.g. $B$) decays into two first generation particles
(e.g. $B \rightarrow D^{**} + W$). Each first generation particle may,
but does not have to, decay into two second generation particles (e.g.
$D^{**} \rightarrow D^*\pi$ and $W \rightarrow l\nu$). Each second
generation particle may, but does not have to, decay into two third
generation particles (e.g. $D^* \rightarrow D\pi$). The process may
be continued.

Each of the $N$ two-body decays ($N=4$ for process in eq. (\ref{DEC}))
is described in
the rest frame of its decaying particle. The masses of all the particles are
considered to be known, therefore, the momenta of the decay
products of each two-body decay are completely specified by the
spherical angles $\theta,\phi$ of the momentum of the first decay
product. The problem is to find the joint decay angular distribution
$W(\theta_1, \phi_1,\ldots,\theta_N, \phi_N)$. This distribution can
be calculated as the trace of the joint spin density matrix of the
$N+1$ final particles:

\begin{equation}
\label{WRH}
W(\theta_1,\ldots,\phi_N) = \sum_{\lambda_1,\ldots,\lambda_{N+1}}
\rho_{\lambda_1,\ldots,\lambda_{N+1}}^{\lambda_1,\ldots,\lambda_{N+1}}
\end{equation}

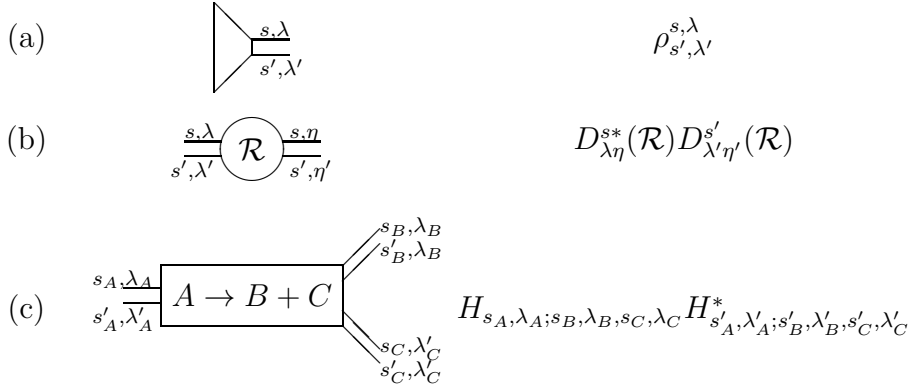
\begin{figure}
\caption{Building blocks for the diagrams\label{r1}}
\begin{center}
\begin{tabular}{ccc}
\raisebox{6mm}{(a)}&
\begin{picture}(10,12)(0,-6)
\thinlines
\put(0,-6){\line(0,1){12}}
\put(0,6){\line(1,-1){5}}
\put(5,1){\line(0,-1){2}}
\put(5,-1){\line(-1,-1){5}}
\put(5,1){\line(1,0){5}}
\put(5,-1){\line(1,0){5}}
\put(6,1.5){\makebox(0,0)[bl]{$\scriptstyle s,\lambda$}}
\put(6,-1.5){\makebox(0,0)[tl]{$\scriptstyle s',
\lambda^{\scriptscriptstyle\prime}$}}
\end{picture}
&
\raisebox{6mm}{$\rho^{s,\lambda}_{s',\lambda'}$}\\
\raisebox{6mm}{(b)}&
\begin{picture}(18,12)(0,-6)
\thinlines
\put(9,0){\circle{8}}
\put(0,-1){\line(1,0){5}}
\put(0,1){\line(1,0){5}}
\put(13,-1){\line(1,0){5}}
\put(13,1){\line(1,0){5}}
\put(9,0){\makebox(0,0){${\cal R}$}}
\put(4,1.5){\makebox(0,0)[br]{$\scriptstyle s,\lambda$}}
\put(4,-1.5){\makebox(0,0)[tr]{$\scriptstyle s',
\lambda^{\scriptscriptstyle\prime}$}}
\put(14,1.5){\makebox(0,0)[bl]{$\scriptstyle s, \eta$}}
\put(14,-1.5){\makebox(0,0)[tl]{$\scriptstyle s',
\eta^{\scriptscriptstyle\prime}$}}
\end{picture}
&
\raisebox{6mm}{$D^{s*}_{\lambda\eta}(\R)D^{s'}_{\lambda'\eta'}(\R)$}\\
\raisebox{9mm}{(c)}&
\begin{picture}(46,24)(-6,-12)
\thinlines
\put(5,-4){\framebox(24,8){$A\rightarrow B+C$}}
\put(0,-1){\line(1,0){5}}
\put(0,1){\line(1,0){5}}
\put(29,4){\line(1,1){5}}
\put(29,2){\line(1,1){5}}
\put(29,-2){\line(1,-1){5}}
\put(29,-4){\line(1,-1){5}}
\put(4,1.5){\makebox(0,0)[br]{$\scriptstyle s_A,\lambda_A$}}
\put(4,-1.5){\makebox(0,0)[tr]{$\scriptstyle s'_A,\lambda'_A$}}
\put(34,8.5){\makebox(0,0)[bl]{$\scriptstyle s_B,\lambda_B$}}
\put(34,7.5){\makebox(0,0)[tl]{$\scriptstyle s'_B,\lambda_B$}}
\put(34,-7.5){\makebox(0,0)[bl]{$\scriptstyle s_C,\lambda'_C$}}
\put(34,-8.5){\makebox(0,0)[tl]{$\scriptstyle s'_C,\lambda'_C$}}
\end{picture}
&
\raisebox{9mm}{$H_{s_A,\lambda_A;s_B,\lambda_B,s_C,\lambda_C}
H^*_{s'_A,\lambda'_A;s'_B,\lambda'_B,s'_C,\lambda'_C}$}
\end{tabular}
\end{center}
\end{figure}

Let us denote the spin density matrix for a single particle by
$\rho^\lambda_{\lambda'}$.  The indices $\lambda,\lambda'$ usually
denote helicities. Sometimes, however, helicity is not sufficient to
define the spin state of a particle. An important
example is the following: a virtual $W$-boson
with helicity zero can be either a spin one (vector) particle or a
spin zero (time like) particle. In such cases it is understood that
each index contains besides the value of helicity also the value of
spin. This, of course, requires a suitable notation. For the virtual
$W$ boson one uses the subscript which is the helicity $\lambda =
\pm1, 0$ for the vector particle and $t$ for the time like particle.
In general, we interpret each index as a pair: $(s,\lambda)$ where $s$
denotes the spin and $\lambda$ the helicity of the particle. If the
particle spin is well defined we will omit the spin specification.
For further use we ascribe to each element
$\rho^\lambda_{\lambda'}$ of the spin density matrix of the parent
particle the diagram shown in Fig \ref{r1}a.

As a result of each two-body decay, the joint spin density matrix of the system
acquires two more indices. Let us take the decay $A
\rightarrow B + C$  for definiteness . The description of this process is
particularly simple
when the spin reference frames shown in Fig. \ref{r2} are used. For this choice
of
frames the axes $z_A$ and $z_B$ are parallel to each other and to the momentum
of particle $B$, as seen from the rest frame of particle $A$ . The
axis $z_C$ is antiparallel to the axes $z_A, z_B$ and parallel to the
momentum of particle $C$ as seen from the rest frame of particle $A$.
The choice of the other axes is described below.

\begin{figure}
\caption{
\label{r2}
Spin reference frames for the decay $A\rightarrow B+C$}
\vspace{6cm}
\end{figure}

Further we shall call the frames like the one shown in Fig \ref{r2} for
particle $A$ decay frames. The decay frame for a particle is defined
using the momenta of next generation particles. Frames like the ones
shown in Fig. \ref{r2} for particles $B$ and $C$ will be called production
frames. They are defined using the momenta of particles from the
preceding generation. Each particle has a production frame, though  the
production frame may be related to a collision
process rather than to a decay for
the parent particle. Each particle that decays in a given
process has a decay frame as well. The rotation which
converts the production frame of particle $i$ into the decay frame of
the same particle will be denoted $R_i$. This rotation can be
parameterized by the three Euler angles $\phi, \theta, \psi$. According
to our construction $\theta$ and $\phi$ are the spherical angles of
the momentum of the first decay product of particle $i$ in the
production frame of particle $i$. We choose the $x$ axis in the decay
frame of particle $i$ so that the third Euler angle $\psi$ is zero.
Geometrically, this means that for particle $i$ the
$z_{production}$-axis, the $z_{decay}$-axis and the $x_{decay}$-axis
are all in one plane. The orientation of the $x_{decay}$-axis is such
that going in the plane from the positive $z_{production}$-semiaxis
past the positive $z_{decay}$-semiaxis one encounters first the
positive $x_{decay}$-semiaxis. The $y_{decay}$-axis is chosen so as to
make the $x_{decay},y_{decay},z_{decay}$ frame right-handed. The
$y_{production}$-axes for the decay products of particle $i$ are
chosen parallel to the $y_{decay}$-axis of particle $i$ as shown in
Fig. \ref{r2}. The $x_{production}$-axes for the decay products are chosen so
as to make the corresponding production frames right-handed. Note that
this makes the $x_{production}$ axis for particle $B$ antiparallel to
the $x_{production}$ axis for particle $C$.

The joint spin density matrix for particles $B$ and $C$ originating from the
two-body decay of particle $A$ is

\begin{equation}
\label{RHO}
\rho^{s_B,\lambda_B,s_C,\lambda_C}_{s'_B,\lambda'_B,s'_C,\lambda'_C} =
\sum_{s_A,\lambda_A,s'_A,\lambda'_A}
\rho^{s_A,\lambda_A}_{s'_A,\lambda'_A}
H_{s_A,\lambda_A;s_B,\lambda_B,s_C,\lambda_C}
H^*_{s'_A,\lambda'_A;s'_B,\lambda'_B,s'_C,\lambda'_C}
\end{equation}

The respective reference frames are defined
in Fig. \ref{r2}, i.e. the spin density matrix for $A$ is in the decay
frame of $A$ and the spin density matrices for $B$ and $C$ are  in the
production frames for $B$ and $C$.  $H_{s,\lambda;\ldots}$ are the helicity
amplitudes where
$s_A$, $\lambda_A$, $s_B$, $\lambda_B$ and $s_C$, $\lambda_C$ are the
spins and helicities of $A$, $B$ and $C$, respectively. The conservation of the
$z$-component of angular momentum implies that the
 helicities of the particles are related as
follows

\begin{equation}
\label{HEL}
\lambda_A = \lambda_B - \lambda_C \hspace{1cm}\mbox{and}\hspace{1cm}
\lambda'_A = \lambda'_B - \lambda'_C
\end{equation}

Relation (\ref{RHO}) corresponds to the diagram shown in Fig \ref{r3}a, where
we have
used the building block introduced in Fig \ref{r1}c. Summations over
indices (helicities and spins) corresponding to the lines without free
ends are understood.
\begin{figure}
\caption{\label{r3}
Diagrams for the decay $A\rightarrow B+C$.
\newline \mbox{\hspace{1cm}}
a) Starting from the decay frame
of particle $A$.
\newline \mbox{\hspace{1cm}}
b) Starting from the production frame of particle $A$.}
\begin{center}
\begin{tabular}{cc}
\begin{picture}(54,36)(0,-12)
\put(0,-6){\begin{picture}(10,12)(0,-6)
\thinlines
\put(0,-6){\line(0,1){12}}
\put(0,6){\line(1,-1){5}}
\put(5,1){\line(0,-1){2}}
\put(5,-1){\line(-1,-1){5}}
\put(5,1){\line(1,0){5}}
\put(5,-1){\line(1,0){5}}
\put(0,0){\makebox(0,0)[l]{$A$}}
\end{picture}}
\put(10,-12){\begin{picture}(36,24)(0,-12)
\thinlines
\put(5,-4){\framebox(24,8){}}
\put(0,-1){\line(1,0){5}}
\put(0,1){\line(1,0){5}}
\put(29,4){\line(1,1){5}}
\put(29,2){\line(1,1){5}}
\put(29,-2){\line(1,-1){5}}
\put(29,-4){\line(1,-1){5}}
\put(34,8.5){\makebox(0,0)[bl]{$\scriptstyle s_B,\lambda_B$}}
\put(34,7.5){\makebox(0,0)[tl]{$\scriptstyle s'_B,\lambda'_B$}}
\put(34,-7.5){\makebox(0,0)[bl]{$\scriptstyle s_C,\lambda_C$}}
\put(34,-8.5){\makebox(0,0)[tl]{$\scriptstyle s'_C,\lambda_C$}}
\put(6,0){\makebox(0,0)[l]{$A\rightarrow B+C$}}
\end{picture}}
\end{picture}
&
\begin{picture}(64,36)(0,-12)
\put(0,-6){\begin{picture}(10,12)(0,-6)
\thinlines
\put(0,-6){\line(0,1){12}}
\put(0,6){\line(1,-1){5}}
\put(5,1){\line(0,-1){2}}
\put(5,-1){\line(-1,-1){5}}
\put(5,1){\line(1,0){5}}
\put(5,-1){\line(1,0){5}}
\put(0,0){\makebox(0,0)[l]{$A$}}
\end{picture}}
\put(10,-6){\begin{picture}(18,12)(0,-6)
\thinlines
\put(9,0){\circle{8}}
\put(0,-1){\line(1,0){5}}
\put(0,1){\line(1,0){5}}
\put(13,-1){\line(1,0){5}}
\put(13,1){\line(1,0){5}}
\put(9,0){\makebox(0,0){${\cal R_A}$}}
\end{picture}}
\put(28,-12){\begin{picture}(36,24)(0,-12)
\thinlines
\put(5,-4){\framebox(24,8){}}
\put(0,-1){\line(1,0){5}}
\put(0,1){\line(1,0){5}}
\put(29,4){\line(1,1){5}}
\put(29,2){\line(1,1){5}}
\put(29,-2){\line(1,-1){5}}
\put(29,-4){\line(1,-1){5}}
\put(34,8.5){\makebox(0,0)[bl]{$\scriptstyle s_B,\lambda_B$}}
\put(34,7.5){\makebox(0,0)[tl]{$\scriptstyle s'_B,\lambda'_B$}}
\put(34,-7.5){\makebox(0,0)[bl]{$\scriptstyle s_C,\lambda_C$}}
\put(34,-8.5){\makebox(0,0)[tl]{$\scriptstyle s'_C,\lambda'_C$}}
\put(6,0){\makebox(0,0)[l]{$A\rightarrow B+C$}}
\end{picture}}
\end{picture}\\
(a)&(b)
\end{tabular}
\end{center}
\end{figure}
In practice the spin density matrix of particle
$A$ is usually known in the production frame. Then in order to use
formula (\ref{RHO}) one must first rotate the spin density matrix of
particle $A$ according to the rule

\begin{equation}
\label{RRO}
\overline\rho^{s,\lambda_A}_{s',\lambda'_A} =
\sum_{\lambda,\lambda'}\rho^{s,\lambda}_{s',\lambda'}
D^{s\,,*}_{\lambda,\lambda_A}(R_A) D^{s'}_{\lambda',\lambda'_A}(R_A)
\end{equation}
where $\rho$ and $\overline\rho$ denote the spin density matrices of particle
$A$ in the production frame and in the decay frame respectively. The diagram
for the calculation of the joint spin density matrix of particles $B$
and $C$ in their production frames  is shown in Fig \ref{r3}b where one
starts from the spin density matrix of particle $A$ in its production
frame. It includes the new building block shown in Fig \ref{r1}b which
corresponds to the rotation of the spin density matrix of particle $A$
from the production frame to the decay frame.

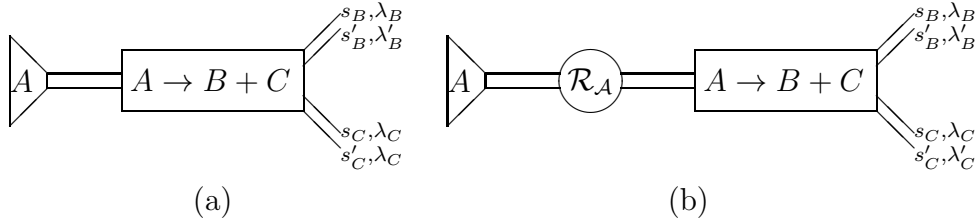
\begin{figure}
\caption{
\label{r4}
Diagram for the angular distribution of particle B from the decay
$A\rightarrow B+C$.}
\begin{center}
\begin{picture}(64,16)(0,-6)
\put(0,-6){\begin{picture}(10,12)(0,-6)
\thinlines
\put(0,-6){\line(0,1){12}}
\put(0,6){\line(1,-1){5}}
\put(5,1){\line(0,-1){2}}
\put(5,-1){\line(-1,-1){5}}
\put(5,1){\line(1,0){5}}
\put(5,-1){\line(1,0){5}}
\put(0,0){\makebox(0,0)[l]{$A$}}
\end{picture}}
\put(10,-6){\begin{picture}(18,12)(0,-6)
\thinlines
\put(9,0){\circle{8}}
\put(0,-1){\line(1,0){5}}
\put(0,1){\line(1,0){5}}
\put(13,-1){\line(1,0){5}}
\put(13,1){\line(1,0){5}}
\put(9,0){\makebox(0,0){${\cal R_A}$}}
\end{picture}}
\put(28,-12){\begin{picture}(38,24)(0,-12)
\thinlines
\put(5,-4){\framebox(24,8){}}
\put(0,-1){\line(1,0){5}}
\put(0,1){\line(1,0){5}}
\put(29,4){\line(1,0){5}}
\put(29,2){\line(1,0){5}}
\put(29,-2){\line(1,0){5}}
\put(29,-4){\line(1,0){5}}
\put(6,0){\makebox(0,0)[l]{$A\rightarrow B+C$}}
\put(34,-3){\oval(2,2)[r]}
\put(34,3){\oval(2,2)[r]}
\end{picture}}
\end{picture}
\end{center}
\end{figure}

As an example let us write down the angular distribution for particle $B$
starting from the spin density matrix of particle $A$ in its production frame.
The diagram shown in Fig. \ref{r4} corresponds to the formula

%CHANGE
\begin{eqnarray}
\label{WBC}
W(\theta,\phi) =
\sum
%_{s,\lambda,s',\lambda',s_B,\lambda_B,s'_B,\lambda'_B,s_C,\lambda_C,
%s'_C,\lambda'_C}
\rho^{s,\lambda}_{s',\lambda'}
D^{s*}_{\lambda,\lambda_A}(\phi,\theta,0)
D^{s'}_{\lambda',\lambda_A}(\phi,\theta,0)* \nonumber \\
H_{s,\lambda_A;\lambda_B,\lambda_C}
H^*_{s',\lambda_A;\lambda_B,\lambda_C}
\end{eqnarray}
where we sum  over all spin and helicity indices and assume that the final
decay
particles have  well defined spins.
Here $\lambda_A$ is defined by relations (\ref{HEL}). We
shall see in the next section how this formula can be simplified.

\section{Statistical tensors and simplified diagrams}

\begin{figure}
\caption{
\label{r5}
Simplified building blocks for diagrams.}
\begin{center}
\begin{tabular}{ccc}
\raisebox{4mm}{(a)}&
\begin{picture}(12,12)(0,-6)
\thinlines
\put(0,-6){\line(0,1){12}}
\put(0,6){\line(1,-1){6}}
\put(6,0){\line(-1,-1){6}}
\put(6,0){\line(1,0){6}}
\put(7,0.5){\makebox(0,0)[lb]{$\scriptstyle JM$}}
\end{picture}
&\raisebox{4mm}{$T^J_M$}
\\
\raisebox{4mm}{(b)}&
\begin{picture}(22,12)(0,-6)
\thinlines
\put(11,0){\circle{8}}
\put(0,0){\line(1,0){7}}
\put(15,0){\line(1,0){7}}
\put(11,0){\makebox(0,0){${\cal R}$}}
\put(16,0.5){\makebox(0,0)[lb]{$\scriptstyle JM'$}}
\put(1,0.5){\makebox(0,0)[lb]{$\scriptstyle JM$}}
\end{picture}
&\raisebox{4mm}{$D^J_{MM'}$}
\\
\raisebox{6mm}{(c)}&
\begin{picture}(46,14)(0,-7)
\thinlines
\put(8,-4){\framebox(24,8){$A\rightarrow B+C$}}
\put(0,0){\line(1,0){8}}
\put(32,4){\line(2,1){8}}
\put(32,-4){\line(2,-1){8}}
\put(0,0.5){\makebox(0,0)[lb]{$\scriptstyle J_AM_A$}}
\put(44,6){\makebox(0,0)[rt]{$\scriptstyle J_BM_B$}}
\put(44,-6){\makebox(0,0)[rb]{$\scriptstyle J_CM_C$}}
\end{picture}
&\raisebox{4mm}{$F^{J_AJ_BJ_C}_{M_AM_BM_C}$}
\end{tabular}
\end{center}
\end{figure}
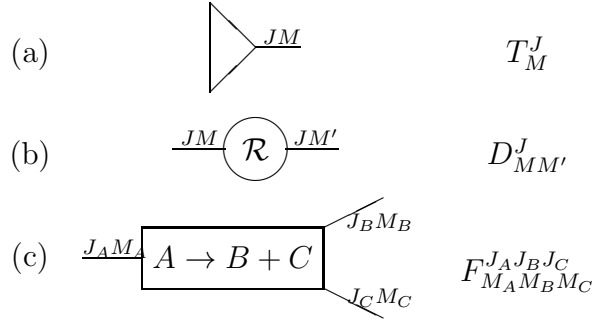

We propose to replace the elements of the spin density matrices used in the
preceding section by the components of the statistical tensors

\begin{equation}
\label{TJM}
T^J_M (s,s') = (-1)^M\sqrt{\frac{2J+1}{2s+1}}\sum_{\lambda,\lambda'}
\rho^{s,\lambda}_{s',\lambda'}<s',\lambda';J,-M|s,\lambda>
\end{equation}
(cf. \cite{BKK} and references contained there). Here $<\ldots|>$ are
Clebsch-Gordan
coefficients and the summation is over the values of $\lambda,\lambda'$
corresponding to non vanishing Clebsch-Gordan coefficients i.e. $-s \leq
\lambda
\leq s$,  $-s' \leq \lambda' \leq s'$ and $\lambda'-\lambda=M$ . With this
replacement the building
blocks of the diagrams are as shown in Fig. \ref{r5}.
Note that now each rotation
introduces only one $D$-function, because formula (\ref{RRO}) is replaced by
(cf. \cite{KZ1} and references contained there)

\begin{equation}
\overline{T}^J_M (s,s') = \sum_{M'=-J}^J{}T^J_{M'}(s,s')D^J_{M'M}(R)
\end{equation}

The coefficients $F^{J_1 J_2 J_3}_{M_1 M_2 M_3}$ from Fig \ref{r5}b are
calculated
according to the formula

%CHANGEIT
\begin{eqnarray}
\label{FJM}
F^{J_1 J_2 J_3}_{M_1 M_2 M_3} &=& \sqrt{\frac{(2J_1+1)(2J_2+1)(2J_3+1)}
{(2s_A+1)(2s_B+1)(2s_C+1)}} \sum_{\lambda_B  \lambda'_B \lambda_C \lambda'_C}
\nonumber \\
& & <s'_A,\lambda'_A;J_1,-M_1|s_A,\lambda_A>
<s'_B,\lambda'_B;J_2,-M_2|s_B,\lambda_B>* \nonumber \\
& & <s'_C,\lambda'_C;J_3,-M_3|s_C,\lambda_C>
H_{s_A,\lambda_A;s_B,\lambda_B,s_C,\lambda_C}
H^*_{s'_A,\lambda'_A;s'_B,\lambda'_B,s'_C,\lambda'_C}
\end{eqnarray}
Here $\lambda_A$ and $\lambda'_A$ are given by the
constraints (\ref{HEL}). The dependence of the coefficients
$F^{J_1 J_2 J_3}_{M_1 M_2 M_3}$ on the decay and on the spins
$s_A,\ldots,s'_C$ has not been  written out. Formula (\ref{FJM})
implies

\begin{equation}
\label{MMM}
M_1 = M_2 - M_3
\end{equation}
and the identity

\begin{eqnarray}
\label{SYM}
F^{J_1 J_2 J_3}_{M_1 M_2 M_3}(s_A,s'_A,s_B,s'_B,s_C,s'_C) =
(-1)^{s_A+s_B+s_C-s'_A-s'_B-s'_C}*  \nonumber \\
F^{J_1 J_2 J_3*}_{-M_1 -M_2 -M_3}(s'_A,s_A,s'_B,s_B,s'_C,s_C)
\end{eqnarray}
which guarantees that the decay angular distribution  is real as it
must be.  Taking the trace over a pair of indices of the joint spin
density matrix is equivalent to coupling the two helicity indices to
$J=0$ according to (\ref{TJM}), then multiplying the result by
$\sqrt{2s + 1}$ and finally summing over all the possible spins $s$ of the
outgoing particle.
Denoting the
summation over a pair of indices by a corner without an outgoing line,
 diagrammatically one has the identity as shown in Fig. \ref{r6}a.

\begin{figure}
\caption{
\label{r6}
Two usefull identities in the diagrammatic form.}
\begin{center}
\begin{tabular}{cc}
\begin{picture}(5,5)
\put(0,0){\line(1,0){5}}
\put(5,0){\line(0,1){5}}
\end{picture}
=$\sum\limits_s\sqrt{2s+1}$
\begin{picture}(16,5)
\put(0,0){\line(1,0){5}}
\put(5,0){\line(0,1){5}}
\put(5,0){\line(2,-1){5}}
\put(9,-2){\makebox(0,0)[lb]{$\scriptstyle J=0$}}
\end{picture}
&
\begin{picture}(34,6)(0,-2)
\put(0,0){\line(1,0){10}}
\put(10,-3){\framebox(20,6){}}
\put(0,.5){\makebox(0,0)[lb]{$\scriptstyle J,M=0$}}
\end{picture}
$=\sqrt{\frac{2J+1}{4\pi}}F(J;s,s')$
\\
(a)&(b)
\end{tabular}
\end{center}
\end{figure}

As a simple example let us reconsider the decay $A \rightarrow B+C$. The
corresponding diagram is shown in Fig. \ref{r7} and replaces the diagram shown
in Fig.
\ref{r4}. The formula for the angular distribution now reads

\begin{equation}
\label{WTJ}
W(\theta,\phi) = \sum_{J, M, s, s'}T^J_M(s, s')F(J;s,s')Y^{J*}_M(\theta,\phi)
\end{equation}
where we have used the well-known identity relating the rotation functions to
spherical harmonics:

\begin{equation}
D^J_{M 0}(\phi,\theta,\psi) = \sqrt{\frac{4\pi}{2J + 1}}Y^{J*}_M(\theta,\phi)
\end{equation}
and the definition of the coefficient \cite{KZ2}

\begin{eqnarray}
F(J;s,s') = N(-1)^{s-s'}\sqrt{\frac{4\pi}{2s+1}}\sum_{\lambda
\lambda'} \nonumber \\
<s,\lambda-\lambda';J,0|s',\lambda-\lambda'>
H_{s,\lambda-\lambda';\lambda,\lambda'}
H^*_{s',\lambda-\lambda';\lambda,\lambda'}
\end{eqnarray}
where $N$ is a suitable normalizing factor. The general identity between a box
with one ingoing line and the coefficient $F(J)$ is shown in Fig \ref{r6}b.
Let us add that the coefficients $F(J)$ can also be defined and
evaluated for more than two body decays (cf. \cite{BKK}, \cite {KZ3}
and references given there).  Thus our analysis can be easily extended
to cascades, where one or more of the final decays is into more than
two particles.

\begin{figure}
\caption{
\label{r7}
Simplified diagram for the decay $A\rightarrow B+C$.}
\begin{center}
\begin{picture}(60,12)(0,-6)
\put(0,-6){\line(0,1){12}}
\put(0,6){\line(1,-1){6}}
\put(6,0){\line(-1,-1){6}}
\put(6,0){\line(1,0){4}}
\put(14,0){\circle{8}}
\put(18,0){\line(1,0){4}}
\put(22,-3){\framebox(20,6){}}
\put(14,0){\makebox(0,0){$\R_A$}}
\end{picture}
\end{center}
\end{figure}

Let us note that a circle in a diagram yields a factor $Y^{J*}_M$ whenever the
index $M$ of the line to the right of the circle equals zero. Because of
identity (\ref{MMM}) this is always the case for the circles preceding the
final decays. Formula (\ref{WTJ}) is clearly simpler than formula (\ref{WBC}).
Moreover, it implies the relation

\begin{equation}
\label{MOM}
a(J,M) \equiv \sum_{s,s'}T^J_M(s,s')F(J;s,s') =
\int{}W(\theta,\phi)Y^J_M(\theta,\phi)d\Omega
\end{equation}
which shows that each number $a(J,M)$ can be obtained from the
experimental angular distribution $W(\theta,\phi)$ and that the set of numbers
$a(J,M)$ contains all the information obtainable from the joint
angular decay distribution.

One could perhaps object that the statistical tensors are less familiar than
the spin density matrices. However, only the statistical tensors for the parent
particle occur explicitly in our formulae. In practice, the parent particle has
either spin zero or spin one-half. In the former case there is only
the $J=0$ statistical tensor

\begin{equation}
T_0^0 = 1
\end{equation}
In the latter, the allowed values for $J$ are $J = 0,1$ and the
necessary  components of the statistical tensor are

\begin{eqnarray}
T_0^0 &=& \frac{1}{\sqrt{2}} \\
T^1_0 &=& \frac{1}{\sqrt{2}}(\rho_{\frac{1}{2} \frac{1}{2}} -
\rho_{-\frac{1}{2} -\frac{1}{2}}) \\
T^1_{\pm{}1} &=& \mp\rho_{\mp{}\frac{1}{2}\pm{}\frac{1}{2}}
\end{eqnarray}
The relation of the components of the statistical tensor to the
elements of the spin density matrices can be seen to be
very simple.
In the following section we apply our algorithm to the decays in eq.
(\ref{DEC}).

\section{Joint angular distributions for the decay cascade
$B \rightarrow
D^{**}( \rightarrow D^*(\rightarrow D\pi) + \pi) + W(\rightarrow l\nu)$}

\begin{figure}
\caption{
\label{r8}
Simplified diagram for the cascade in eq. (1).}
\begin{center}
\begin{picture}(100,46)(0,-16)
\put(0,-6){\line(0,1){12}}
\put(0,6){\line(1,-1){6}}
\put(6,0){\line(-1,-1){6}}
\put(0,0){\makebox(0,0)[l]{$B$}}
\put(6,0){\line(1,0){4}}
\put(10,-4){\framebox(24,8){$B\rightarrow D^{**}W$ }}
\put(34,4){\line(2,1){4}}
\put(46,10){\line(2,1){4}}
\put(34,-4){\line(2,-1){4}}
\put(46,-10){\line(2,-1){4}}
\put(42,8){\circle{8}}
\put(42,8){\makebox(0,0){$\R_{D^{**}}$}}
\put(42,-8){\circle{8}}
\put(42,-8){\makebox(0,0){$\R_{W}$}}
\put(50,-16){\framebox(24,8){$W\rightarrow l\nu$}}
\put(50,8){\framebox(24,8){$D^{**}\rightarrow D^*\pi$}}
\put(74,16){\line(1,0){4}}
\put(86,16){\line(1,0){4}}
\put(82,16){\circle{8}}
\put(82,16){\makebox(0,0){$\R_{D^*}$}}
\put(90,12){\framebox(24,8){$D^*\rightarrow D\pi$}}
\end{picture}
\end{center}
\end{figure}
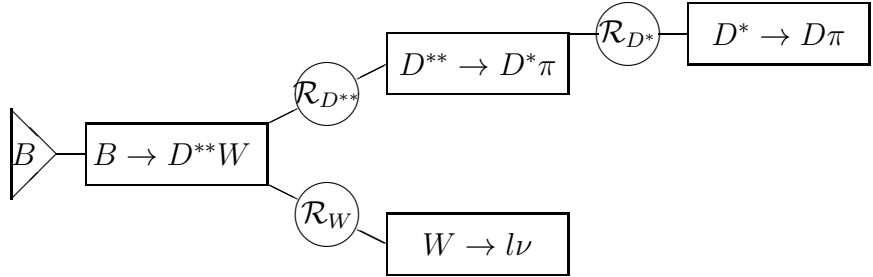

The diagrammatic representation of the
joint decay distributions for the decays in eq. (\ref{DEC}) is  shown
in Fig \ref{r8}. The corresponding decay formula reads

{\samepage
\begin{eqnarray}
\label{WD1}
W(\theta_l,\phi_l,\theta_{D*},\phi_{D^*},\theta_D,\phi_D) =
\sum_{J_3 = 0,2} \sum_{J_2 = 0}^2 \sum_{J_1 = 0}^{2 s_{D^{**}}}
\sum_{M = 0}^{J_{12}}
\sum_{M_1=-J_{13}}^{J_{13}}{}^{\mbox{\hspace{-2mm}}\prime}
\nonumber \\
Re \left[ a(J_1, J_2, J_3, M, M_1) D^{J_1}_{M M_1}(\phi_{D^*}, \theta_{D^*},
0)\right.
\left. Y^{J_2*}_M(\theta_l,\phi_l)
Y^{J_3*}_{M_1}(\theta_D, \phi_D) \right] \nonumber \\
(2 - \delta_{M 0} \delta_{M_1 0})
\end{eqnarray}}
\noindent where the identity (\ref{SYM}) has been used. $J_{i j}$ denotes the
smaller of the two numbers $J_i$ and $J_j$. The prime in the last
sum means that  only  nonnegative $M_1$ are to be included
for $M = 0$. The coefficients $a(J_1,J_2,J_3,M,M_1)$ are given by the
formula

\begin{equation}
\label{ASM}
a(J_1, J_2, J_3, M, M_1) = F_{D^*}(J_3) F^{J_1 J_3 0}_{M_1 M_1 0} \sum_{s_W
s_{W'}=0,1} F^{0 J_1 J_2}_{0 M M}(s_W, s'_W) F_W(J_2; s_W, s'_W)
\end{equation}

Each of the coefficients $a(J_1,J_2,J_3,M,M_1)$ is independently measurable
because formula (\ref{WD1}) implies the relation

\begin{eqnarray}
\label{MSM}
a(J_1,\ldots,M_1) = \frac{2J_1 + 1}{4\pi}\int d\Omega_l d\Omega_{D^*} d\Omega_D
W(\theta_l,\ldots,\phi_D)* \nonumber \\
\left[D^{J_1*}_{MM_1}(\phi_{D^*}, \theta_{D^*}, 0)
Y^{J_2}_M(\theta_l,\phi_l) Y^{J_3}_{M_1}(\theta_D, \phi_D)\right]
\end{eqnarray}
The integral is just the average of the expression in square brackets over the
experimental joint decay distribution $W(\theta_l,\ldots)$. In
principle, the averaging should be performed at fixed $W$-boson mass
$\sqrt{q^2}$. In practice, one would probably use some model for the
$q^2$-dependence of the helicity amplitudes in order to be able to use
all the data simultaneously. The coefficients $F^{0 J_1 J_2}_{0 M
M}(s_W,s'_W)$ correspond to the decay $B \rightarrow D^{**}+W$ and the
coefficients $F^{J_1 J_3 0}_{M_1 M_1 0}$ to the decay $D^{**}
\rightarrow D^*\pi$. The two possible spins of the $W$-boson  $s_W,
s'_W = 0,1$ have been taken into account. For each of the two decays
the spin state of one particle ($\lambda = \pm{}1,0,t$ for the $W$ and
$\lambda = \pm1,0$ for the $D^*$) and angular momentum conservation
fix the spin states of the other two particles participating in the
corresponding two-body decay.  Therefore, we shall use the notation
$H_\lambda$ and $h_\lambda$ for the helicity amplitudes of the two decays,
respectively. We normalize the angular distribution
$W(\theta_l,\ldots)$ so as to get as simple formulae as possible.
In some of the following expressions overall factors will be dropped.
Note also
that because of the equality between the subscript of $Y^{J_2}_M(\theta_l,
\phi_l)$ and
the first subscript of $D^{J_1}_{MM_1}(\phi_{D^*},\theta_{D^*},0)$,
the azimuthal angles $\phi_l$ and
$\phi_{D^*}$  always occur in the combination $\phi_l +
\phi_{D^*} = \pi - \chi$ where $\chi$ is the angle between the half
planes bounded by the $z_{B decay}$ axis and containing the momenta of
the lepton and of the $D^*$, respectively. Thus  the
distribution $W(\theta_l,\ldots)$ is effectively given in term of five
variables instead of six.

The coefficients $F_i(J)$ have been reviewed and listed in ref. \cite{BKK}.
The nonzero values needed for the present case are

\begin{eqnarray}
\label{FPO}
F_{D^*}(0) &=& \sqrt{\frac{3}{4\pi}}\\
F_{D^*}(2) &=& -\sqrt{\frac{3}{10\pi}}\\
F_W(0;0,0) &=& \sqrt{\frac{1}{4\pi}}\frac{3\epsilon}{1+\epsilon}\\
F_W(1;1,0) &=& -F_W(1;0,1) =
-\sqrt{\frac{3}{4\pi}}\frac{\epsilon}{1+\epsilon}\\
F_W(0;1,1) &=& \sqrt{\frac{3}{4\pi}}\\
F_W(1;1,1) &=& \pm\sqrt{\frac{3}{8\pi}}\frac{1}{1+\epsilon}\label{pm}\\
\label{FKO}
F_W(2;1,1) &=& \sqrt{\frac{3}{40\pi}}\frac{1-2\epsilon}{1+\epsilon}
\end{eqnarray}
where    lepton mass effects ($m_l\neq0$) are taken into account by
introducing the ratio ( $q^2$ is the invariant mass squared of the
lepton pair)

\begin{equation}
\epsilon = \frac{m_l^2}{2 q^2}
\end{equation}
The two possible  decay cases $W^-\rightarrow l^-+\bar\nu_l$ and
$W^+\rightarrow l^++\nu_l$ are both included in eqs. (\ref{pm}) where only
$F_W(1;1,1)$ in (\ref{ASM}) is affected ($\pm$ for $W^\pm$). It is seen that
the terms with $J_3=1$ vanish in the expansion (\ref{WD1}).

Let us consider the coefficients $F^{J_1 J_3 0}_{M_1 M_1 0}$ for the decays
$D^{**}(1^+) \rightarrow D^*\pi$. Parity conservation implies $h_1 =
h_{-1}$. Therefore, the dynamics of the decay $D^{**}(1^+) \rightarrow D^*\pi$
introduces  two free
parameters (the overall normalization is not counted) into the joint angular
distribution, which may be
combined into one complex parameter $\eta$, if we put

\begin{equation}
\label{ETA}
h_0 = \eta h_1
\end{equation}

According to HQET $\eta=1$ for $D^{**}(1^+, j=\frac{1}{2}) \rightarrow
D^*\pi$ and $\eta = -2$ for $D^{**}(1^+, j=\frac{3}{2}) \rightarrow D^*\pi$.
{}From  (\ref{FJM}) one has (see e.g. [13,14]) %CHANGEIT

\begin{eqnarray}
F^{0 0 0}_{0 0 0} &=&\frac{1}{3}(2+|\eta|^2) \\
F^{2 0 0}_{0 0 0} = F^{0 2 0}_{0 0 0} &=& \frac{\sqrt{2}}{3}(1-|\eta|^2) \\
F^{1 2 0}_{1 1 0} &=& Im(\eta)\\
F^{2 2 0}_{0 0 0} &=&\frac{1}{3}(1+2|\eta|^2) \\
F^{2 2 0}_{1 1 0} &=& Re(\eta)\\
F^{2 2 0}_{2 2 0} &=& 1
\end{eqnarray}
where the overall factor $|h_1|^2$ has been dropped. It can be seen
that $(J_1,M_1) = (1,0)$ is forbidden.

For the decay $D^{**}(2^+) \rightarrow D^*\pi$ parity conservation implies
$h_1 = -h_{-1}$ and $h_0=0$. For the coefficients $F^{J_1 J_3 0}_{M_1 M_1 0}$
one finds

\begin{eqnarray}
F^{0 0 0}_{0 0 0} &=& \frac{2}{\sqrt{15}}\\
F^{2 0 0}_{0 0 0} &=& -\sqrt{\frac{2}{21}}\\
F^{0 2 0}_{0 0 0} &=& \sqrt{\frac{2}{15}}\\
F^{2 2 0}_{0 0 0} &=& -\sqrt{\frac{1}{21}}\\
F^{2 2 0}_{\pm2 \pm2 0} &=& -\sqrt{\frac{3}{7}}\\
F^{4 0 0}_{0 0 0} &=& -4\sqrt{\frac{2}{105}}\\
F^{4 2 0}_{0 0 0} &=& -4\sqrt{\frac{1}{105}}\\
F^{4 2 0}_{\pm2 \pm2 0} &=& \frac{2}{\sqrt{7}}
\end{eqnarray}
where again the overall factor $|h_1|^2$ has been dropped. Thus the decay
$D^{**}(2^+) \rightarrow D^* + \pi$ introduces no free parameters into the
joint angular distribution. One also notices that there are no contributions
with $J_1 = 1,3$ or $M_1 = \pm1$.

Finally, we must consider the coefficients $F^{0 J_1 J_2}_{0 M M}(s_W, s'_W)$.
Let us begin with the decay $B\rightarrow D^{**}(1^+)+W$. For $s_W = s'_W =1$
we find

\begin{eqnarray}
\label{F1P}
F^{0 0 0}_{0 0 0}&=& \frac{1}{3}(|H_{-1}|^2 + |H_0|^2 + |H_1|^2)\\
F^{0 1 0}_{0 0 0} &=& F^{0 0 1}_{0 0 0} = \frac{1}{\sqrt{6}}(|H_1|^2| -
|H_{-1}|^2)\\
F^{0 1 1}_{0 0 0} &=& \frac{1}{2}(|H_{-1}|^2 + |H_1|^2)\\
F^{0 1 1}_{0 1 1} &=& \frac{1}{2}(H_{-1}H_0^* + H_0 H_1^*)\\
F^{0 0 2}_{0 0 0} = F^{0 2 0}_{0 0 0} &=& \frac{1}{3\sqrt{2}}(|H_{-1}|^2 -
2|H_0|^2 + |H_1|^2)\\
F^{0 1 2}_{0 0 0} = F^{0 2 1}_{0 0 0} &=& \frac{1}{2\sqrt{3}}(|H_1|^2 -
|H_{-1}|^2)\\
F^{0 1 2}_{0 1 1} = F^{0 2 1}_{0 1 1} &=& \frac{1}{2}(-H_{-1}H_0^* + H_0
H_1^*)\\
F^{0 2 2}_{0 0 0} &=& \frac{1}{6}(|H_{-1}|^2 + 4|H_0|^2 + |H_1|^2)\\
F^{0 2 2}_{0 1 1} &=& \frac{1}{2}(H_{-1}H_0^* + H_0 H_1^*)\\
\label{F1K}
F^{0 2 2}_{0 2 2} &=& H_{-1}H^*_1
\end{eqnarray}
while for $s_W = s'_W=0$:

\begin{eqnarray}
F^{0 0 0}_{0 0 0} &=& \frac{1}{\sqrt{3}}|H_t|^2\\
F^{0 2 0}_{0 0 0} &=& -\sqrt{\frac{2}{3}}|H_t|^2
\end{eqnarray}
For the interference terms $F^{0 J 1}_{0 M M}(0,1) = -F^{0 J 1*}_{0 -M
-M}(1, 0)$. For $s_W=1$, $s'_W=0$:

\begin{eqnarray}
F^{0 0 1}_{0 0 0} &=& \frac{1}{\sqrt{3}}H_0 H^*_t\\
F^{0 1 1}_{0 \pm1 \pm1} &=& \pm\frac{1}{\sqrt{2}}H_{\mp1} H^*_t\\
F^{0 2 1}_{0 0 0} &=& -\sqrt{\frac{2}{3}}H_0 H^*_t\\
F^{0 2 1}_{0 \pm1 \pm1} &=& -\sqrt{\frac{1}{2}}H_{\mp1} H^*_t
\end{eqnarray}
Substituting all these coefficients into formula (\ref{WD1}) one obtains the
distribution $W(\theta_l,\ldots,\phi_D)$ as a sum of 43 terms. Each of the 43
coefficients $a(J_1, J_2, J_3, M, M_1)$ is experimentally measurable according
to formula (\ref{MOM}). These coefficients, however, are all functions of the
same helicity amplitudes $H_1, H_{-1}, H_0, H_t, h_1, h_0$. Since the overall
normalization and the overall phase of the amplitudes for each decay drop out
from the formula for the joint angular distribution the number of real
dynamical parameters that are  to be extracted from the data is eight.
The problem can be seen to be  grossly overdetermined.
Nevertheless, each measured coefficient $a(J_1, J_2, J_3, M, M_1)$ can
be used to reduce the error of  measuring the helicity
amplitudes.

For the decay $ B \rightarrow D^{**}(2^+)+W$ and $s_W = s'_W =1$:

\begin{eqnarray}
F^{0 0 0}_{0 0 0} &=& \frac{1}{\sqrt{15}}(|H_{-1}|^2 + |H_0|^2+|H_1|^2)\\
F^{0 1 0}_{0 0 0} &=& \frac{1}{\sqrt{30}}(|H_1|^2 - |H_{-1}|^2)\\
F^{0 0 1}_{0 0 0} &=& \frac{1}{\sqrt{10}}(|H_1|^2  - |H_{-1}|^2)\\
F^{0 2 0}_{0 0 0} &=& -\frac{1}{\sqrt{42}}(|H_1|^2  + 2|H_0|^2 + |H_{-1}|^2)\\
F^{0 0 2}_{0 0 0} &=& \frac{1}{\sqrt{30}}(|H_1|^2  - 2|H_0|^2 + |H_{-1}|^2)\\
F^{0 2 1}_{0 0 0} &=& -\frac{1}{2\sqrt{7}}(|H_1|^2  - |H_{-1}|^2)\\
F^{0 2 1}_{0 1 1} &=& \frac{1}{2\sqrt{7}}(H_0 H_1^*  - H_{-1}H_0^*)\\
F^{0 2 2}_{0 0 0} &=& -\frac{1}{2\sqrt{21}}(|H_1|^2  - 4|H_0|^2 + |H_{-1}|^2)\\
F^{0 2 2}_{0 1 1} &=& \frac{1}{2\sqrt{7}}(H_0 H_1^*  + H_{-1}H_0^*)\\
F^{0 2 2}_{0 2 2} &=& \sqrt{\frac{3}{7}}H_{-1} H_1^*\\
F^{0 4 0}_{0 0 0} &=& -\sqrt{\frac{2}{105}}(2|H_1|^2  - 3|H_0|^2 +
2|H_{-1}|^2)\\
F^{0 4 1}_{0 0 0} &=& -\frac{2}{\sqrt{35}}(|H_1|^2  - |H_{-1}|^2)\\
F^{0 4 1}_{0 1 1} &=& -\sqrt{\frac{3}{14}}(H_0 H_1^*  - H_{-1}H_0^*)\\
F^{0 4 2}_{0 0 0} &=& -2\sqrt{\frac{1}{105}}(|H_1|^2  + 3|H_0|^2 +
|H_{-1}|^2)\\
F^{0 4 2}_{0 1 1} &=& -\sqrt{\frac{3}{14}}(H_0 H_1^*  + H_{-1}H_0^*)\\
F^{0 4 2}_{0 2 2} &=& -\frac{2}{\sqrt{7}}H_{-1}H_1^*
\end{eqnarray}
while for  $s_W=s'_W=0$ one has

\begin{eqnarray}
F^{0 0 0}_{0 0 0} &=& \sqrt{\frac{1}{5}}|H_t|^2\\
F^{0 2 0}_{0 0 0} &=& -\sqrt{\frac{2}{7}}|H_t|^2\\
F^{0 4 0}_{0 0 0} &=& 3\sqrt{\frac{2}{35}}|H_t|^2
\end{eqnarray}
Finally, for the interference terms $F^{0 J 1}_{0 M M}(0,1)=-F^{0 J 1*}_{0 -M
-M}(1,0)$ and for $s_W=1, s'_W=0$:

\begin{eqnarray}
F^{0 0 1}_{0 0 0} &=& \sqrt{\frac{1}{5}}H_0 H_t^*\\
F^{0 2 1}_{0 0 0} &=& -\sqrt{\frac{2}{7}}H_0 H_t^*\\
F^{0 2 1}_{0 \pm1 \pm1} &=& -\sqrt{\frac{1}{14}}H_{\mp1} H_t^*\\
F^{0 4 1}_{0 0 0} &=& \frac{6}{\sqrt{70}}H_0 H_t^*\\
F^{0 4 1}_{0 \pm1 \pm1} &=& \sqrt{\frac{3}{7}}H_{\mp1}
H_t^* \end{eqnarray}
In this case the expansion (\ref{WD1}) consists of 48
terms. The 48 coefficients $a(J_1, J_2, J_3, M, M_1)$ depend on six
real parameters corresponding to the helicity amplitudes $H_1, H_{-1},
H_0, H_t$.

In order to reduce the number of free parameters in the joint angular
distribution to below eight in the $D^{**}(1^+)$ case and to below six in
the $D^{**}(2^+)$ case, it is necessary to know something about the
dynamics of the decay processes involved. The HQET predictions
for $B \rightarrow D^{**}(J^P) + W$ (cf.
\cite{IW2,KZ0,BKT,KKP})  are given by %CHANGEIT

\begin{eqnarray}
\label{H1P}
H_t(1^+) &\sim& H_0(2^+)\sim\sqrt{\frac{\omega+1}{q^2}}(M_B - M_D)\\
H_0(1^+) &\sim& H_t(2^+)\sim\sqrt{\frac{\omega-1}{q^2}}(M_B + M_D)\\
\label{H1K}
H_{\pm1}(1^+,j=\frac{1}{2}) &\sim& -2H_{\pm1}(1^+, j=\frac{3}{2}) \sim
\pm\sqrt{\omega+1} -\sqrt{\omega - 1}\\
H_{\pm1}(2^+) &\sim& \frac{\sqrt{3}}{2}(\sqrt{\omega +1} \mp \sqrt{\omega -
1})
\end{eqnarray}
where $\omega$ is the velocity transfer variable $\omega=v_1v_2$ as
usual.

In these formulae the symbol $\sim$ means equal up to a factor
independent of $\lambda$. Factors dependent on $J^P$ and $j$ have been
dropped. Thus, these formulae should not be used to compare different
decays. For given $J^P$ and $j$, however, they yield all the
amplitudes $H_\lambda(J^P, j)$ up to a constant factor and consequently are
sufficient to calculate  all
the coefficients $F^{0 J_1 J_2}_{0 M M }$ from the formulae given in
this section . Combined with the values of the parameter $\eta$ given
previously, they give absolute and complete predictions for the joint
angular distributions of all the cascade decays in eq. (\ref{DEC}).

\section{Applications}
Once one has measured  the coefficients
$a(J_1,J_2,J_3,M,M_1)$ it is possible to determine (up to a common
complex factor) all the helicity amplitudes $H_\lambda$ and $h_\lambda$. In
fact the problem is grossly overdetermined, so that many strategies
and many consistency checks are possible. In practice the choice will
be dictated by the  available data. For the $D^{**}(2^+)$ it is
necessary to know at least six ratios of the coefficients
$a(J_1,J_2,J_3,M,M_1)$ in order to find all the amplitudes $H_\lambda$. This
reduces to four ratios when the lepton mass is neglected and to three
ratios, if
the amplitudes are assumed to be relatively real (sign ambiguities
remain).  Neglecting both the lepton mass and the relative phases, one
is left with two real numbers to be determined. In the $D^{**}(1^+)$
case one may be interested either in the amplitudes $H_\lambda$ or in the
amplitudes $h_\lambda$. We show below how the two problems can be
separated from one another.

Let us consider the ratios of the coefficients $a(J_1,J_2,J_3,M,M_1)$ with
common values of $J_1,J_2$ and $M$. In such ratios the amplitudes $H_\lambda$
cancel. If the lepton-side information is not used, i.e. taking
$J_2=M=0$, we find for the $D^{**}(1^+)$ cases:

\begin{eqnarray}
\frac{a(0,0,2,0,0)}{a(0,0,0,0,0)} =
\frac{2}{\sqrt{5}}\frac{|\eta|^2-1}{|\eta|^2+2}\\
\frac{a(2,0,0,0,0)}{a(2,0,2,0,2)} = \frac{\sqrt{5}}{3}(|\eta|^2-1)\\
\frac{a(2,0,2,0,0)}{a(2,0,2,0,2)} = \frac{1}{3}(1+2|\eta|^2)\\
\frac{a(2,0,2,0,1)}{a(2,0,2,0,2)} = Re(\eta)
\end{eqnarray}

These ratios yield the complex parameter $\eta$ up to the sign of
$Im(\eta)$.  Since one expects theoretically that $Im(\eta)$ is close
to zero, the determination of this sign may be difficult. In principle
it can be deduced from the sign of the coefficient $a(1,1,2,0,1)$.
The vanishing of $Im(\eta)$, on the other hand, is easy to test,
because it implies the vanishing of five coefficients.  Integrating
the decay distribution over both $d\Omega_l$ and $d\Omega_D$, i.e.
considering only the decays $D^{**}(1^+) \rightarrow D^* + \pi$, one
is left with two coefficients: $a(0,0,0,0,0)$ and $a(2,0,0,0,0)$.
Their ratio depends both on the amplitudes $H_\lambda$ and on the amplitudes
$h_\lambda$ and, of course, is not sufficient to find either set. This
is the problem pointed out for the amplitudes $h_\lambda$ in ref.
\cite{LWI}.  Our solution is to look at the joint decay distribution
of $D^{**} \rightarrow D^*(\rightarrow D + \pi) + \pi$.  Contrary to
ref.  \cite{LWI}, we make no assumptions about the spin density matrix
of the $D^{**}$. The HQET predictions are obtained by substituting
$\eta=1$ and $\eta=-2$ in the case $j=\frac{1}{2}$ and
$j=\frac{3}{2}$, respectively.

In order to find the amplitudes $H_\lambda$ for the $D^{**}(1^+)$ case, we
propose
to look at the decay distributions integrated over $d\Omega_D$, i.e. at  the
coefficients with $J_3 = M_1 =0$, and to compare pairs of coefficients with the
same  values of $J_1$. Then the dependence on the amplitudes $h_\lambda$ drops
out. We find seven equations for the six unknown parameters:

\begin{eqnarray}
\frac{a(0,1,0,0,0)}{a(0,0,0,0,0)} &=& \frac{\sqrt{3}}{2(1+\epsilon)}
\frac{\pm(|H_1|^2-|H_{-1}|^2) -
4\epsilon{}Re(H_0H^*_t)}{|H_1|^2+|H_{-1}|^2+|H_{0t}|^2}\\
\frac{a(0,2,0,0,0)}{a(0,0,0,0,0)} &=&
\frac{1}{2\sqrt{5}}\frac{1-2\epsilon}{1+\epsilon}
\frac{|H_1|^2-2|H_0|^2+|H_{-1}|^2}{|H_1|^2+|H_{-1}|^2+|H_{0t}|^2}\\
\frac{a(2,2,0,1,0)}{a(2,2,0,0,0)} &=&
3\frac{H_{-1}H^*_0+H_0H^*_1}{|H_1|^2+4|H_0|^2+|H_{-1}|^2}\\
\frac{a(2,2,0,2,0)}{a(2,2,0,0,0)} &=&
6\frac{H_{-1}H^*_1}{|H_1|^2+4|H_0|^2+|H_{-1}|^2}\\
\frac{a(2,0,0,0,0)}{a(2,2,0,0,0)} &=& 2\sqrt{5}\frac{1+\epsilon}{1-2\epsilon}
\frac{|H_1|^2-2|H_{0t}|^2+|H_{-1}|^2}{|H_1|^2+4|H_0|^2+|H_{-1}|^2}\\
\frac{a(2,1,0,0,0)}{a(2,2,0,0,0)}
&=& \frac{\sqrt{15}}{1-2\epsilon}
\frac{\pm(|H_1|^2-|H_{-1}|^2)+8\epsilon{}Re(H_0H^*_t)}
{|H_1|^2+4|H_0|^2+|H_{-1}|^2}\\
\frac{a(2,1,0,1,0)}{a(2,2,0,0,0)} &=& \frac{3\sqrt{5}}{1-2\epsilon}
\frac{\mp(H_{-1}H^*_0-H_0H^*_1)+2\epsilon(H_{-1}H^*_t+H_tH^*_1)}
{|H_1|^2+4|H_0|^2+|H_{-1}|^2}
\end{eqnarray}
where we have used the abbreviation

\begin{equation}
|H_{0t}|^2 = |H_0|^2 + \frac{3\epsilon}{1+\epsilon}|H_t|^2
\end{equation}
{}From these relations one can find all the amplitudes $H_\lambda$ (up to a
constant factor) and make one consistency checks. The HQET predictions
can be obtained by substituting the amplitudes $H_\lambda(1^+)$ from
formulae (\ref{H1P}) - (\ref{H1K}).

The analysis for the $D^{**}(2^+)$ case is much simpler, because there is no
problem with the amplitudes $h_\lambda$. The integration over $d\Omega_l$ and
$d\Omega_D$ leaves, however, also in this case too few coefficients
($a(0,0,0,0,0), a(2,0,0,0,0)$ and $a(4,0,0,0,0)$) and the study of some joint
decay distributions is necessary. From the three coefficients one can only find
the ratio

\begin{equation}
R = \frac{|H_{0t}|^2}{|H_1|^2 + |H_{-1}|^2}
\end{equation}
and make one consistency check.

In order to experimentally determine the amplitudes $h_\lambda$ one
can also use other decays. For the decay $B \rightarrow
D^{**}(\rightarrow D^*(\rightarrow D\pi)+ \pi)+\pi$ the joint angular
distribution can be formally obtained from the formulae given in
Section 4 by putting $H_t=1, H_{\pm1}=H_0=0,F_W(0;0,0)=1$ and
omitting the angular $\theta_l,\phi_l$-dependence. The number of
parameters $a(J_1,J_2,J_3,M,M_1)$ left is six for $D^{**}(1^+)$ and
eight for $D^{**}(2^+)$. For the $D^{**}(2^+)$ case angular momentum
and parity  conservation
(for the $D^{**}$ decay) are sufficient to
make an absolute prediction. For the $D^{**}(1^+)$ case the six
parameters depend on the complex parameter $\eta$. The same argument
applies when the first generation pion is replaced by any other spin
zero meson. Another example is the nonleptonic decay $B \rightarrow
D^{**}(\rightarrow D^*(\rightarrow D\pi)+\pi)+\rho(\rightarrow
\pi\pi)$. The necessary replacements in  the formulae
of  Section 5  are the following:
(i) $H_t=0$, (ii) omission of the arguments $s_W, s'_W$,
(iii) replacement of $\theta_l,\phi_l$ by $\theta_\pi, \phi_\pi$
corresponding to one of the decay products of the $\rho$ and
(iv) insertion of $F_\rho(0) = \sqrt{\frac{3}{4\pi}}$, $F_\rho(1)=0$
and $F_\rho(2)=\sqrt{\frac{3}{10\pi}}$ for $F_W(0;1,1)$, $F_W(1;1,1)$
and $F_W(2;1,1)$. The problem is similar to the problem
for the decays in eq. (\ref{DEC}) when the lepton mass is neglected.
The number of measurable parameters, however, is reduced to 25 for the
$D^{**}(1^+)$ and to 32 for the $D^{**}(2^+)$. The $\rho$-meson can be
replaced by any spin one meson decaying into two spin zero particles.

\section{Conclusions}

The joint angular distribution for each of the decays
in eq. (\ref{DEC}) depends on more than 40 independently measurable
parameters. The coefficients $a(J_1,J_2,J_3,M,M_1)$ are a convenient complete
set and can also be used to express other parameters. For instance, the lepton
energy $E_l$ in the $B$ rest frame is a linear function of
cos$\theta_l$. Consequently, the average $<E_l>$
can be calculated as a linear function of the coefficient
$a(0,1,0,0,0)$. The average $<E_l^2>$ depends, moreover,
linearly on the coefficient $a(0, 2, 0, 0, 0)$, while all the averages
$<E_l^n>$ for $n~>~2$ can be expressed as linear functions of
these two. The decays of  $D^{**}$-s into longitudinally polarized
$D^*$-s studied in \cite{LWI} correspond to a determination of the
coefficients $a(J_1, 0, J_3, 0, 0)$.

The coefficients $a(J_1,J_2,J_3,M,M_1)$, though independently measurable, are
strongly correlated as a result of angular momentum conservation. Thus, when
testing models
by comparison with data
it is preferable to compare with  the predictions for the
helicity amplitudes and not for the coefficients $a(J_1,J_2,J_3,M,M_1)$.
Otherwise it is easy to misinterpret implications of angular momentum
conservation as successful predictions of one's model. In Section 4 we have
given the predictions of HQET, which would be very interesting to check. It is
also important to test the relative reality of the helicity amplitudes. The
helicity amplitudes can, of course, be replaced by form factors (cf e.g.
\cite{BKK}).

{}From experiment one would like to know in each case the coefficient
$a(0,0,0,0,0)$, (which defines the overall normalization), and as many other
coefficients as possible, because each new measured coefficient
reduces the experimental uncertainties. It is important to realize
that single-step decay distributions contain little information. Thus
it is crucial to look at the joint decay distributions of at least
pairs of particles. As discussed in Section 5, the joint two-particle
decay distributions are sufficient to find all the helicity
amplitudes. Nevertheless, the study of the full triple decay
distributions can significantly reduce the errors in the determination
of the helicity amplitudes.

\end{document}